\documentclass[11pt]{article}
\usepackage[T1]{fontenc}
\usepackage[utf8]{inputenc}
\usepackage[a4paper]{geometry}
\usepackage[active]{srcltx}
\usepackage{color}
\usepackage{textcomp}
\usepackage{amsmath}
\usepackage{amssymb}
\usepackage{esint}
\usepackage{todonotes}

\providecommand{\Lt}{{\tt L}}
\renewcommand{\Lt}{{\tt L}}
\providecommand{\Wt}{{\tt W}}
\renewcommand{\Wt}{{\tt W}}
\providecommand{\Xt}{{\tt X}}
\renewcommand{\Xt}{{\tt X}}
\providecommand{\Yt}{{\tt Y}}
\renewcommand{\Yt}{{\tt Y}}

\makeatletter

\usepackage{textcomp}

\pdfoutput=1 

\usepackage{jheppub}

\usepackage{etoolbox}
    
    \patchcmd{\maketitle}{\@fpheader}{}{}{}

\usepackage{amsfonts}

\usepackage{bbm}

\title{Higher Spin Black Holes with Soft Hair}

\author[]{Daniel Grumiller$^{1}$,}
\author[]{Alfredo P\'{e}rez$^{2}$,}
\author[]{Stefan Prohazka$^{1}$,}
\author[]{David Tempo$^{2}$}
\author[]{and Ricardo Troncoso$^{2}$}

\affiliation[]{$^{1}$Institute for Theoretical Physics, TU Wien, Wiedner Hauptstrasse 8--10/136, A-1040 Vienna, Austria, }
\affiliation[]{$^{2}$Centro de Estudios Cient\'{i}ficos (CECs), Av. Arturo Prat 514, Valdivia,
Chile.}

\emailAdd{grumil@hep.itp.tuwien.ac.at}
\emailAdd{aperez@cecs.cl}
\emailAdd{prohazka@hep.itp.tuwien.ac.at}
\emailAdd{tempo@cecs.cl}
\emailAdd{troncoso@cecs.cl}

\preprint{CECS-PHY-16/03, TUW-16-13}

\abstract{We construct a new set of boundary conditions for higher spin gravity, inspired by a recent ``soft Heisenberg hair''-proposal for General Relativity on three-dimensional Anti-de~Sitter. The asymptotic symmetry algebra consists of a set of affine $\hat u(1)$ current algebras. Its associated canonical charges generate higher spin soft hair. We focus first on the spin-3 case and then extend some of our main results to spin-$N$, many of which resemble the spin-2 results: the generators of the asymptotic $W_3$ algebra naturally emerge from composite operators of the $\hat u(1)$ charges through a twisted Sugawara construction; our boundary conditions ensure regularity of the Euclidean solutions space independently of the values of the charges; solutions, which we call ``higher spin black flowers'', are stationary but not necessarily spherically symmetric. Finally, we derive the entropy of higher spin black flowers, and find that for the branch that is continuously connected to the BTZ black hole, it depends only on the affine purely gravitational zero modes. Using our map to $W$-algebra currents we recover well-known expressions for higher spin entropy. We also address higher spin black flowers in the metric formalism and achieve full consistency with previous results.}

\makeatother

\begin{document}
\maketitle \flushbottom

\section{Introduction}
\label{Introduction}

It has recently been shown that General Relativity with negative cosmological
constant in three spacetime dimensions admits a new class of stationary
black holes that are not necessarily spherically symmetric and do
not fulfill the Brown--Henneaux boundary conditions \cite{Afshar:2016wfy}.
A consistent set of boundary conditions that accommodates this class
of solutions was then constructed, from which it can be seen that
the generic black flower configuration is endowed with an infinite
number of affine $\hat{u}\left(1\right)$ charges that commute with
the total Hamiltonian. These charges then correspond to soft hair
in the sense of Hawking, Perry and Strominger \cite{Hawking:2016msc}.
It was also shown in \cite{Afshar:2016wfy} that the precise levels
of the $\hat{u}\left(1\right)$ generators are such that the Virasoro
algebra with the Brown--Henneaux central extension \cite{Brown:1986nw} naturally emerges
as a composite one through a twisted Sugawara construction. An additional
remarkable feature of the boundary conditions aforementioned is that
the spectrum, given by the space of solutions that fulfill them, is
such that regularity of the configurations at the horizon holds independently
of the global charges.

One of our main goals is to extend these results to the case of gravity
on AdS$_{3}$ coupled to bosonic higher spin fields \cite{Blencowe:1988gj,Bergshoeff:1989ns,Vasiliev:1995dn}.
In this case, it is known that the infinite tower of fields can be
consistently truncated so as to describe a finite number of interacting
nonpropagating fields of spin $s=2$, $3$, $\cdots$, $N$, see e.g.
\cite{Henneaux:2010xg,Campoleoni:2010zq}. The theory can then be
generically described in terms of the difference of two Chern--Simons
actions for independent gauge fields $A^{\pm}$ that take values in
$sl\left(N,\mathbb{R}\right)$, so that the action reads
\begin{equation}
I=I_{CS}\left[A^{+}\right]-I_{CS}\left[A^{-}\right]\;,\label{eq:HS-Action}
\end{equation}
with
\begin{equation}
I_{\mathrm{CS}}[A]=\frac{k_{N}}{4\pi}\int\text{tr}\left(A\wedge dA+\frac{2}{3}A\wedge A\wedge A\right)\;,\label{CS-Action}
\end{equation}
where $\text{tr}\left(\cdots\right)$ stands for the trace in the
fundamental representation of $sl\left(N,\mathbb{R}\right)$ (see
Appendix \ref{Appendix:A}). The level in \eqref{CS-Action} relates
to the Newton constant and the AdS radius according to $k_{N}=\frac{k}{2\epsilon_{N}}=\frac{\ell}{8G\epsilon_{N}}$,
whose normalization is determined by $\epsilon_{N}=\frac{N(N^{2}-1)}{12}$.

The gauge fields are related to a suitable generalization of the dreibein
and the spin connection, defined through 
\begin{equation}
A^{\pm}=\omega\pm\frac{e}{\ell}
\end{equation}
and hence, the spacetime metric and the higher spin fields can be
reconstructed from
\begin{equation}
g_{\mu\nu}=\frac{1}{\epsilon_{N}}\text{tr}(e_{\mu}e_{\nu}) \qquad\qquad \Phi_{\mu_{1}\dots\mu_{s}}=\frac{1}{\epsilon_{N}^{(s)}}\text{tr}(e_{(\mu_{1}}\dots e_{\mu_{s})}) \label{MetricAndSpin-S-Fields}
\end{equation}
being manifestly invariant under an extension of the local Lorentz
group described by the diagonal subgroup of $SL\left(N,\mathbb{R}\right)\otimes SL\left(N,\mathbb{R}\right)$.
Diffeomorphisms and higher spin transformations are then related to
the remaining gauge symmetries.
The parentheses around the subscripts denote symmetrization and the spin-3 field in our $sl(3,\mathbb{R})$ calculations is normalized such that $\epsilon_{3}^{(3)}=3!$.

This paper is organized as follows. In section \ref{Asymptotic-structure} we present our boundary conditions for the spin-3 case in diagonal gauge and derive the asymptotic symmetry algebra as well as higher spin soft hair. In section \ref{Mapping} we map our boundary conditions into highest weight gauge variables and exhibit the twisted Sugawara-like construction. In section \ref{higher spin black hole} we focus on higher spin black holes with soft hair, which we call ``higher spin black flowers''. We derive their entropy and find that it depends only on our purely gravitational zero modes; in fact, it is linear in them, exactly as in General Relativity. However, if expressed in terms of $W$-algebra currents we recover well-known expressions that involve non-polynomial expressions of spin-2 and spin-3 charges. We further demonstrate that our boundary conditions ensure that all states in our theory are compatible with the regularity conditions. In section \ref{sec:higher-spin-black} we address higher spin black flowers in the metric formalism and achieve full consistency with previous results. In section \ref{GenericN} we generalize some of the main results to spin-$N$, based on $sl(N,\mathbb{R})$ Chern--Simons theories (with principally embedded $sl(2,\mathbb{R})$). In section \ref{sec:discussion} we conclude with a discussion and future research directions.

\section{Asymptotic structure}
\label{Asymptotic-structure}

The asymptotic structure of AdS gravity coupled to higher spin fields
in three-dimensional spacetimes was investigated in \cite{Henneaux:2010xg,Campoleoni:2010zq},
where it was shown that the asymptotic symmetries are spanned by two
chiral copies of $W$ algebras (see also \cite{Gaberdiel:2011wb,Campoleoni:2011hg,Campoleoni:2014tfa}).
In order to accommodate the different higher spin black hole solutions
in \cite{Gutperle:2011kf,Castro:2011fm}, and \cite{Henneaux:2013dra,Bunster:2014mua},
the asymptotic behaviour has to be extended so as to incorporate chemical
potentials associated to the global charges. In this sense, alternative
proposals have been constructed. The one in \cite{Gutperle:2011kf,Ammon:2011nk}
successfully accommodates the black hole solution with higher spin
fields of \cite{Gutperle:2011kf}, while the set of boundary conditions
in \cite{Henneaux:2013dra,Bunster:2014mua} does for the higher spin black holes described
therein. It is worth pointing out that the asymptotic symmetries of
both sets are different.

Here we construct an inequivalent set of boundary conditions, which
reduces to the one recently introduced in \cite{Afshar:2016wfy} when
the higher spin fields are switched off. The asymptotic behaviour
of the $sl\left(3,\mathbb{R}\right)$ gauge fields is proposed to
be given by 
\begin{equation}
A^{\pm}=b_{\pm}^{-1}\left(d+a^{\pm}\right)b_{\pm}\label{Amn}
\end{equation}
so that the dependence on the radial coordinate is completely contained
in the group elements
\begin{equation}
b_{\pm}=\exp\left(\pm\frac{1}{\ell\zeta^{\pm}}\Lt_{1}\right)\cdot\exp\left(\pm\frac{\rho}{2}\Lt_{-1}\right)\;.\label{b-null}
\end{equation}
The auxiliary connection reads
\begin{equation}
a^{\pm}=\left(\pm\mathcal{J}^{\pm}\ d\varphi+\zeta^{\pm}\ dt\right) \Lt_{0} + \left(\pm\mathcal{J}_{\left(3\right)}^{\pm}\ d\varphi+\zeta_{\left(3\right)}^{\pm}\ dt\right)\Wt_{0} \label{amn}
\end{equation}
where $\Lt_{i}, \Wt_{n}$, with $i=-1,0,1$, and $n=-2,-1,0,1,2$, span
the $sl\left(3,\mathbb{R}\right)$ algebra. Following~\cite{Henneaux:2013dra},
it can be seen that $\mathcal{J}^{\pm}$ and $\mathcal{J}_{\left(3\right)}^{\pm}$
stand for arbitrary functions of (advanced) time and the angular coordinate
that correspond to the dynamical fields, while $\zeta^{\pm}$ and
$\zeta_{\left(3\right)}^{\pm}$ describe their associated Lagrange
multipliers that can be assumed to be fixed at the boundary without
variation ($\delta\zeta^{\pm}=\delta\zeta_{\left(3\right)}^{\pm}=0$). We shall refer to $\zeta^\pm$, $\zeta^\pm_{(3)}$ as chemical potentials.

The field equations, implying the local flatness of the gauge fields,
then reduce to 
\begin{equation}
\dot{\mathcal{J}}^{\pm}=\pm\zeta^{\prime} \qquad\qquad \dot{\mathcal{J}}_{\left(3\right)}^{\pm}=\pm\zeta_{\left(3\right)}^{\prime}\;,\label{FE-DG}
\end{equation}
where dot and prime denote derivatives with respect to $t$ and $\varphi$,
respectively.

\subsection{Asymptotic symmetries and the algebra of the canonical generators}

In the canonical approach \cite{Regge:1974zd}, the variation of the conserved charges 
\begin{equation}
Q[\epsilon^{+},\epsilon^{-}]={\cal Q}^{+}[\epsilon^{+}]-{\cal Q}^{-}[\epsilon^{-}]
\end{equation}
associated to gauge symmetries spanned by $\epsilon^{\pm}=\epsilon_{i}^{\pm} \Lt_{i}+\epsilon_{\left(3\right)n}^{\pm} \Wt_{n}$,
that maintain the asymptotic form of the gauge fields, is determined
by
\begin{equation}
\delta{\cal Q}^{\pm}\left[\epsilon^{\pm}\right]=\mp\frac{k}{4\pi}\int d\varphi\left(\eta^{\pm}\delta\mathcal{J}^{\pm}+\frac{4}{3}\eta_{\left(3\right)}^{\pm}\delta\mathcal{J}_{\left(3\right)}^{\pm}\right)\;,
\end{equation}
with $\eta^{\pm}=\epsilon_{0}^{\pm}$, and $\eta_{\left(3\right)}^{\pm}=\epsilon_{\left(3\right)0}^{\pm}$.
According to \eqref{amn}, the asymptotic symmetries fulfill $\delta_{\epsilon^{\pm}}a^{\pm}=d\epsilon^{\pm}+[a^{\pm},\,\epsilon^{\pm}]={\cal O}(\delta a^{\pm})$,
provided that the transformation law of the dynamical fields reads
\begin{equation}
\delta\mathcal{J}^{\pm}=\pm\eta^{\pm\prime} \quad\quad \delta\mathcal{J}_{\left(3\right)}^{\pm}=\pm\eta_{\left(3\right)}^{\pm\prime} \label{TransfLaw JmnJ3mn}
\end{equation}
and the parameters are time-independent ($\dot{\eta}^{\pm}=\dot{\eta}_{\left(3\right)}^{\pm}=0$).
One has to take the quotient over the remaining
components of $\epsilon^{\pm}$, since they just span trivial gauge
transformations that neither appear in the variation of the global
charges nor in the transformation law of the dynamical fields. 

The surface integrals that correspond to the conserved charges associated
to the asymptotic symmetries then readily integrate as
\begin{equation}
{\cal Q}^{\pm}\left[\eta^{\pm},\eta_{\left(3\right)}^{\pm}\right]=\mp\frac{k}{4\pi}\int d\varphi\left(\eta^{\pm}\left(\varphi\right)\mathcal{J}^{\pm}\left(\varphi\right)+\frac{4}{3}\eta_{\left(3\right)}^{\pm}\left(\varphi\right)\mathcal{J}_{\left(3\right)}^{\pm}\left(\varphi\right)\right)\;,\label{Qmn}
\end{equation}
which are manifestly independent of the radial coordinate $\rho$.
Consequently, the boundary could be located at any fixed value $\rho=\rho_{0}$.
Hereafter, we assume that $\rho_{0}\rightarrow\infty$, since this
choice has the clear advantage of making our analysis to cover the
entire spacetime in bulk.

The algebra of the global charges can then be directly obtained from
the computation of their Poisson brackets; or as a shortcut, by virtue
of $\delta_{Y}Q\left[X\right]=\{Q\left[X\right],Q\left[Y\right]\}$,
from the variation of the dynamical fields in \eqref{TransfLaw JmnJ3mn}.
Expanding in Fourier modes
\begin{equation}
\label{eq:Jmodes}
\mathcal{J}^{\pm}\left(\varphi\right)=\frac{2}{k}\sum_{n=-\infty}^{\infty}J_{n}^{\pm}e^{\pm in\varphi} \qquad \qquad \mathcal{J}_{\left(3\right)}^{\pm}\left(\varphi\right)=\frac{3}{2k}\sum_{n=-\infty}^{\infty}J_{n}^{\left(3\right)\pm}e^{\pm in\varphi}
\end{equation}
leads to the asymptotic symmetry algebra which is described by a set
of $\hat{u}\left(1\right)$ currents whose nonvanishing brackets are
given by 
\begin{equation}
i\left\{ J_{n}^{\pm},J_{m}^{\pm}\right\} =\frac{1}{2}kn\delta_{m+n,0} \qquad \qquad i\left\{ J_{n}^{\left(3\right)\pm},J_{m}^{\left(3\right)\pm}\right\} =\frac{2}{3}kn\delta_{m+n,0}\;,\label{algebra3}
\end{equation}
with levels $\frac{1}{2}k$, and $\frac{2}{3}k$, respectively. 

\subsection{(Higher spin) soft hair}
\label{soft hair}

Following the spin-2 construction \cite{Afshar:2016wfy}, we consider now all vacuum descendants $|\psi(q)\rangle$ labelled by a set $q$ of non-negative integers $N^\pm$, $N_{(3)}^\pm$, $n_i^\pm$, $n_i^{(3)\,\pm}$, $m_i^\pm$ and $m_i^{(3)\,\pm}$
  \begin{equation}
\label{eq:angelinajolie}
\big|\psi(q)\rangle = N(q)\prod_{i=1}^{N^\pm}\Big(J^\pm_{-n_i^\pm}\Big)^{m_i^\pm}\prod_{i=1}^{N^\pm_{(3)}}\Big(J^{(3)\,\pm}_{-n_i^{(3)\,\pm}}\Big)^{m_i^{(3)\,\pm}}\big|0\rangle \,.
  \end{equation}
Here $N(q)$ is some normalization constant such that $\langle\psi(q)|\psi(q)\rangle=1$ and the vacuum state is defined through highest weight conditions, $J_n^\pm|0\rangle=J_n^{(3)\,\pm}|0\rangle=0$ for non-negative $n$.

We want to check now if all vacuum descendants $|\psi(q)\rangle$ have the same energy as the vacuum and are thus soft hair (our discussion easily generalizes from soft hair descendants of the vacuum to soft hair descendants of any higher spin black hole state). To this end we consider the surface integral associated with the generator in time, given by
\begin{equation}
H:=Q\left(\partial_{t}\right)=\frac{k}{4\pi}\int d\varphi\left(\zeta^{+}\mathcal{J}^{+}+\zeta^{-}\mathcal{J}^{-}+\frac{4}{3}\zeta_{\left(3\right)}^{+}\mathcal{J}_{\left(3\right)}^{+}+\frac{4}{3}\zeta_{\left(3\right)}^{-}\mathcal{J}_{\left(3\right)}^{-}\right)\;.
\end{equation}
For constant chemical potentials $\zeta^{\pm}$,
$\zeta_{\left(3\right)}^{\pm}$ the field equations \eqref{FE-DG}
imply that the dynamical fields become time-independent, and the total
Hamiltonian reduces to
\begin{equation}
H=\zeta^{+}J_{0}^{+}+\zeta^{-}J_{0}^{-}+\zeta_{\left(3\right)}^{+}J_{0}^{\left(3\right)+}+\zeta_{\left(3\right)}^{-}J_{0}^{\left(3\right)-}\;,
\end{equation}
which clearly commutes with the whole set of asymptotic symmetry generators
spanned by $J_{n}^{\pm}$ and $J_{m}^{\left(3\right)\pm}$. One then
concludes that for an arbitrary fixed value of the total energy, configurations
endowed with different sets of nonvanishing $\hat{u}\left(1\right)$
charges turn out to be inequivalent, because they can not be related
to each other through a pure gauge transformation. Since excitations \eqref{eq:angelinajolie} associated with the generators $J_{n}^{\pm}$, $J_{m}^{\left(3\right)\pm}$ preserve the total energy and cannot be gauged away, they are (higher spin) soft hair in the sense of Hawking, Perry and Strominger \cite{Hawking:2016msc}.

\section{Highest weight gauge and the emergence
of composite $W_{3}$ symmetries}
\label{Mapping}

Quite remarkably, it can be seen that spin-2 and spin-3 charges naturally
emerge as composite currents constructed out from the $\hat{u}(1)$
ones. 
Actually, the full set of generators of the $W_{3}$ algebra
arises from suitable composite operators of the $\hat{u}(1)$ charges
through a twisted Sugawara construction. Here we show this explicitly
through the comparison of the new set of boundary conditions proposed
in the previous section with the ones that accommodate the higher
spin black holes in \cite{Henneaux:2013dra,Bunster:2014mua}, whose asymptotic symmetries
are described by two copies of the $W_{3}$ algebra. In order to carry
out this task it is necessary to express both sets in terms of the
same variables. The asymptotic behaviour
described by \eqref{Amn} and \eqref{amn} is formulated so that the
auxiliary connections $a^{\pm}$ are written in the diagonal gauge,
while the set in \cite{Henneaux:2013dra,Bunster:2014mua} was formulated in the so-called
highest weight gauge. Consequently, what we look for can be unveiled
once the gauge fields in \eqref{Amn} and \eqref{amn} are expressed
in terms of the variables that are naturally adapted to the gauge
fields $\hat{A}^{\pm}$ in the highest weight gauge. 

For a generic choice of Lagrange multipliers, which are still unspecified,
the asymptotic form of the gauge fields in the highest weight gauge
reads \cite{Henneaux:2013dra,Bunster:2014mua}
\begin{equation}
\hat{A}^{\pm}=\hat{b}_{\pm}^{-1}(d+\hat{a}^{\pm})\hat{b}_{\pm}\;,
\end{equation}
where the radial dependence can be captured by the choice $\hat{b}_{\pm}=e^{\pm\rho \Lt_{0}}$,
and 
\begin{align}
\hat{a}_{\varphi}^{\pm} & =\Lt_{\pm1}-\frac{2\pi}{k}\mathcal{L}_{\pm} \Lt_{\mp1} - \frac{\pi}{2k}\mathcal{W}_{\pm}\Wt _{\mp2} \qquad \qquad \hat{a}_{t}^{\pm}=\Lambda^{\pm}\left[\mu_{\pm},\nu_{\pm}\right]\;,\label{amnHW}
\end{align}
with
\begin{align}
\Lambda^{\pm}\left[\mu_{\pm},\nu_{\pm}\right] & =\pm\left[\mu_{\pm}\Lt_{\pm1}+\nu_{\pm}\Wt_{\pm2}\mp\mu_{\pm}^{\prime}\Lt_{0}\mp\nu_{\pm}^{\prime}\Wt_{\pm1}+\frac{1}{2}\left(\mu_{\pm}^{\prime\prime}-\frac{4\pi}{k}\mu_{\pm}\mathcal{L}_{\pm}+\frac{8\pi}{k}\mathcal{W}_{\pm}\nu_{\pm}\right)\Lt_{\mp1}\right.\nonumber \\
 & \quad -\left(\frac{\pi}{2k}\mathcal{W}_{\pm}\mu_{\pm}+\frac{7\pi}{6k}\mathcal{L}_{\pm}^{\prime}\nu_{\pm}^{\prime}+\frac{\pi}{3k}\nu_{\pm}\mathcal{L}_{\pm}^{\prime\prime}+\frac{4\pi}{3\kappa}\mathcal{L}_{\pm}\nu_{\pm}^{\prime\prime}\right.\left.-\frac{4\pi^{2}}{k^{2}}\mathcal{L}_{\pm}^{2}\nu_{\pm}-\frac{1}{24}\nu_{\pm}^{\prime\prime\prime\prime}\right)\Wt_{\mp2}\nonumber \\
 & \quad +\left.\frac{1}{2}\left(\nu_{\pm}^{\prime\prime}-\frac{8\pi}{k}\mathcal{L}_{\pm}\nu_{\pm}\right)\Wt_{0}\mp\frac{1}{6}\left(\nu_{\pm}^{\prime\prime\prime}-\frac{8\pi}{k}\nu_{\pm}\mathcal{L}_{\pm}^{\prime}-\frac{20\pi}{k}\mathcal{L}_{\pm}\nu_{\pm}^{\prime}\right)\Wt_{\mp 1}\right]\ ,\label{Lambda-HW}
\end{align}
where ${\cal L}_{\pm}$, ${\cal W}_{\pm}$ and $\mu_{\pm}$, $\nu_{\pm}$
stand for arbitrary functions of $t,\varphi$.

One then needs to find suitable permissible gauge transformations
spanned by group elements $g_{\pm}$, for which $\hat{a}^{\pm}=g_{\pm}^{-1}\left(d+a^{\pm}\right)g_{\pm}$.
These group elements indeed exist, and they are given by $g_{\pm}=g_{\pm}^{\left(1\right)}g_{\pm}^{\left(2\right)}$,
with
\begin{align}
g_{\pm}^{\left(1\right)} & =\exp\left[x_{\pm} \Lt_{\pm1}+y_{\pm} \Wt_{\pm1}+z_{\pm} \Wt_{\pm2}\right] \nonumber \\
g_{\pm}^{\left(2\right)} & =\exp\left[-\frac{1}{2}{\cal J}^{\pm} \Lt_{\mp1}-\frac{1}{3}{\cal J}_{\left(3\right)}^{\pm} \Wt_{\mp1}\pm\frac{1}{6}\left({\cal J}^{\pm}{\cal J}_{\left(3\right)}^{\pm}+\frac{1}{2}{\cal J}_{\left(3\right)}^{\pm\prime}\right) \Wt_{\mp2}\right]\;,\label{g1mn-g2mn}
\end{align}
where $x_{\pm}$, $y_{\pm}$ and $z_{\pm}$ are arbitrary functions
of $t$, $\varphi$ that fulfill the following conditions:
\begin{align}
x_{\pm}^{\prime} & =1+{\cal J}^{\pm}x_{\pm}+2{\cal J}_{\left(3\right)}^{\pm}y_{\pm} \nonumber \\
y_{\pm}^{\prime} & =2{\cal J}_{\left(3\right)}^{\pm}x_{\pm}+{\cal J}^{\pm}y_{\pm} \label{XYZprime}\\
z_{\pm}^{\prime} & =2{\cal J}^{\pm}z_{\pm}\mp\frac{1}{2}y_{\pm}\;,\nonumber 
\end{align}
with
\begin{align}
\mu_{\pm} & =-x_{\pm}\zeta^{\pm}-2y_{\pm}{\cal \zeta}_{\left(3\right)}^{\pm}\pm\dot{x}_{\pm}\pm\frac{4}{3}\nu_{\pm}{\cal J}_{\left(3\right)}^{\pm} \nonumber \\
\nu_{\pm} & =-2\zeta_{\pm}z_{\pm}\pm\left(x_{\pm}^{2}-y_{\pm}^{2}\right){\cal \zeta}_{\left(3\right)}^{\pm}+\frac{1}{2}\left(y_{\pm}\dot{x}_{\pm}-x_{\pm}\dot{y}_{\pm}\pm2\dot{z}_{\pm}\right)\;.\label{XYZdot}
\end{align}
Consistency of eqs. \eqref{XYZprime} and \eqref{XYZdot} on-shell implies that the Lagrange multipliers in the highest weight gauge,
$\mu_{\pm}$, $\nu_{\pm}$, depend not only on the Lagrange multipliers
in the diagonal gauge, $\zeta^{\pm}$, $\zeta_{\left(3\right)}^{\pm}$,
but also on their corresponding global charges ${\cal J}^{\pm}$,
${\cal J}_{\left(3\right)}^{\pm}$, according to
\begin{align}
\pm\frac{2}{3}{\cal J}_{\left(3\right)}^{\pm}\nu_{\pm}^{\prime}\mp\frac{8}{3}\left({\cal J}^{\pm}{\cal J}_{\left(3\right)}^{\pm}+\frac{1}{2}{\cal J}_{\left(3\right)}^{\pm\prime}\right)\nu_{\pm}+\mu_{\pm}^{\prime}-\mu_{\pm}{\cal J}^{\pm} & =-\zeta^{\pm} \label{eq:MuJ}\\
\frac{1}{2}\nu_{\pm}^{\prime\prime}-\frac{3}{2}{\cal J}_{\pm}\nu_{\pm}^{\prime}+\left(\left({\cal J}^{\pm}\right)^{2}-\frac{4}{3}\left({\cal J}_{\left(3\right)}^{\pm}\right)^{2}-{\cal J}^{\pm\prime}\right)\nu_{\pm}\pm\mu_{\pm}{\cal J}_{\left(3\right)}^{\pm} & =\pm{\cal \zeta}_{\left(3\right)}^{\pm}\;.\label{eq:MuU}
\end{align}
The gauge fields $a^{\pm}$ and $\hat{a}^{\pm}$ are then mapped to
each other provided
\begin{align}
{\cal L}_{\pm} & =\pm\frac{k}{4\pi}\left(\frac{1}{2}\left({\cal J}^{\pm}\right)^{2}+\frac{2}{3}\left({\cal J}_{\left(3\right)}^{\pm}\right)^{2}+{\cal J}^{\pm\prime}\right) \label{eq:MiuraL}\\
{\cal W}_{\pm} & =\mp\frac{k}{6\pi}\left(-\frac{8}{9}\left({\cal J}_{\left(3\right)}^{\pm}\right)^{3}+2\left({\cal J}^{\pm}\right)^{2}{\cal J}_{\left(3\right)}^{\pm}+{\cal J}_{\left(3\right)}^{\pm}{\cal J}^{\pm\prime}+3{\cal J}^{\pm}{\cal J}_{\left(3\right)}^{\pm\prime}+{\cal J}_{\left(3\right)}^{\pm\prime\prime}\right) \label{eq:MiuraW}
\end{align}
from which one recognizes the Miura transformation between the variables,
see e.g. \cite{Bouwknegt:1992wg}.

In sum, our proposal for boundary conditions once expressed in the
highest weight gauge, is such that the Lagrange multipliers $\mu_{\pm}$
and $\nu_{\pm}$ depend on the dynamical variables according to \eqref{eq:MuJ}
and \eqref{eq:MuU}, where $\zeta^{\pm}$ and $\zeta_{\left(3\right)}^{\pm}$
are assumed to be fixed without variation ($\delta\zeta^{\pm}=\delta\zeta_{\left(3\right)}^{\pm}=0$).
Note that the functions ${\cal L}_{\pm}$, ${\cal W}_{\pm}$, that
are naturally defined in the highest weight gauge, depend on the global
charges ${\cal J}^{\pm}$, ${\cal J}_{\left(3\right)}^{\pm}$ as in
eqs. \eqref{eq:MiuraL}, \eqref{eq:MiuraW}. 

Indeed, for a generic choice of Lagrange multipliers in the highest
weight gauge, the field equations read \cite{Bunster:2014mua}
\begin{align}
\dot{\mathcal{L}}_{\pm} & =\pm2\mathcal{L}_{\pm}\mu_{\mathcal{\pm}}^{\prime}\pm\mu_{\pm}\mathcal{L}_{\pm}^{\prime}\mp\frac{k}{4\pi}\mu_{\mathcal{\pm}}^{\prime\prime\prime}\mp2\nu_{\pm}\mathcal{W}_{\pm}^{\prime}\mp3\mathcal{W}_{\pm}\nu_{\pm}^{\prime}\label{eq:LPuntoW3-1}\\
\dot{\mathcal{W}}_{\pm} & =\pm3\mathcal{W}_{\pm}\mu_{\mathcal{\pm}}^{\prime}\pm\mu_{\pm}\mathcal{W}_{\pm}^{\prime}\pm\frac{2}{3}\nu_{\pm}\left(\mathcal{L}_{\pm}^{\prime\prime\prime}-\frac{16\pi}{k}\mathcal{L}_{\pm}^{2\prime}\right)\pm3\left(\mathcal{L}_{\pm}^{\prime\prime}-\frac{64\pi}{9k}\mathcal{L}_{\pm}^{2}\right)\nu_{\pm}^{\prime}\nonumber \\
 &\quad \pm5\nu_{\pm}^{\prime\prime}\mathcal{L}_{\pm}^{\prime}\pm\frac{10}{3}\mathcal{L}_{\pm}\nu_{\pm}^{\prime\prime\prime}\mp\frac{k}{12\pi}\nu_{\pm}^{\left(5\right)}\ ,\label{eq:WPuntoW3}
\end{align}
which by virtue of the definition of our boundary conditions, in eqs.
\eqref{eq:MuJ}, \eqref{eq:MuU} and \eqref{eq:MiuraL}, \eqref{eq:MiuraW},
reduce to the remarkably simple ones, given by $\dot{\mathcal{J}}^{\pm}=\pm\zeta^{\prime}$,
$\dot{\mathcal{J}}_{\left(3\right)}^{\pm}=\pm\zeta_{\left(3\right)}^{\prime}$,
which were directly obtained in the diagonal gauge (see eq. \eqref{FE-DG}).

It is also worth highlighting that eqs. \eqref{eq:MiuraL}, \eqref{eq:MiuraW}
can be regarded as the higher spin gravity version of the twisted
Sugawara construction. In fact, we show now that the currents ${\cal L}_\pm$, ${\cal W}_\pm$ fulfill the $W_3$ algebra. Let us recall that
according to \eqref{TransfLaw JmnJ3mn}, the transformation law of
the dynamical fields under the $\hat{u}\left(1\right)$ asymptotic
symmetries reads $\delta\mathcal{J}^{\pm}=\pm\eta^{\pm\prime}$ ,
$\delta\mathcal{J}_{\left(3\right)}^{\pm}=\pm\eta_{\left(3\right)}^{\pm\prime}$.
Besides, the relationship for the Lagrange multipliers in \eqref{eq:MuJ},
\eqref{eq:MuU} implies that the corresponding one for the parameters
in the highest weight gauge, defined as $\varepsilon_{\pm}$, $\chi_{\pm}$,
with the Lagrange multipliers and their associated charges as formulated
in the diagonal gauge, given by $\eta^{\pm}$, $\eta_{\left(3\right)}^{\pm}$,
and ${\cal J}^{\pm}$, ${\cal J}_{\left(3\right)}^{\pm}$, respectively,
reads
\begin{align}
\pm\frac{2}{3}{\cal J}_{\left(3\right)}^{\pm}\chi_{\pm}^{\prime}\mp\frac{8}{3}\left({\cal J}^{\pm}{\cal J}_{\left(3\right)}^{\pm}+\frac{1}{2}{\cal J}_{\left(3\right)}^{\pm\prime}\right)\chi_{\pm}+\varepsilon_{\pm}^{\prime}-\varepsilon_{\pm}{\cal J}^{\pm} & =-\eta^{\pm} \nonumber \\
\frac{1}{2}\chi_{\pm}^{\prime\prime}-\frac{3}{2}{\cal J}_{\pm}\chi_{\pm}^{\prime}+\left(\left({\cal J}^{\pm}\right)^{2}-\frac{4}{3}\left({\cal J}_{\left(3\right)}^{\pm}\right)^{2}-{\cal J}^{\pm\prime}\right)\chi_{\pm}\pm\varepsilon_{\pm}{\cal J}_{\left(3\right)}^{\pm} & =\pm{\cal \eta}_{\left(3\right)}^{\pm} \label{mmp2} \; .
\end{align}
Therefore, the transformation laws for ${\cal L}_{\pm}$ and ${\cal W}_{\pm}$
can be directly read from \eqref{eq:MiuraL}, \eqref{eq:MiuraW},
which reduce to 
\begin{align}
\delta\mathcal{L}_{\pm} & =\pm2\mathcal{L}_{\pm}\varepsilon_{\mathcal{\pm}}^{\prime}\pm\varepsilon_{\pm}\mathcal{L}_{\pm}^{\prime}\mp\frac{k}{4\pi}\varepsilon_{\mathcal{\pm}}^{\prime\prime\prime}\mp2\chi_{\pm}\mathcal{W}_{\pm}^{\prime}\mp3\mathcal{W}_{\pm}\chi_{\pm}^{\prime} \\
\delta\mathcal{W}_{\pm} & =\pm3\mathcal{W}_{\pm}\varepsilon_{\mathcal{\pm}}^{\prime}\pm\varepsilon_{\pm}\mathcal{W}_{\pm}^{\prime}\pm\frac{2}{3}\chi_{\pm}\left(\mathcal{L}_{\pm}^{\prime\prime\prime}-\frac{16\pi}{k}\mathcal{L}_{\pm}^{2\prime}\right)\pm3\left(\mathcal{L}_{\pm}^{\prime\prime}-\frac{64\pi}{9k}\mathcal{L}_{\pm}^{2}\right)\chi_{\pm}^{\prime}\nonumber \\
 &\quad \pm 5 \chi_{\pm}^{\prime\prime}\mathcal{L}_{\pm}^{\prime}\pm\frac{10}{3}\mathcal{L}_{\pm}\chi_{\pm}^{\prime\prime\prime}\mp\frac{k}{12\pi}\chi_{\pm}^{\left(5\right)}\ .
\end{align}
It is then apparent that ${\cal L}_{\pm}$ and ${\cal W}_{\pm}$ turn
out to be composite anomalous spin-2 and spin-3 currents, respectively.
In other words, the asymptotic $W_{3}$ algebra obtained in \cite{Henneaux:2013dra,Bunster:2014mua}
for a different set of boundary conditions, being defined through
requiring the Lagrange multipliers in the highest weight gauge to
be fixed without variation ($\delta\mu_{\pm}=\delta\nu_{\pm}=0$),
is recovered as a composite one that emerges from the $\hat{u}\left(1\right)$
currents.

Despite of the fact that the spin-2 and spin-3 currents
${\cal L}_{\pm}$, ${\cal W}_{\pm}$ fulfill the $W_{3}$ algebra,
their associated global charges generate the $\hat{u}\left(1\right)$
current algebras discussed in section \eqref{Asymptotic-structure}.
This is so because, by virtue of eqs. in \eqref{mmp2}
and \eqref{eq:MiuraL}, \eqref{eq:MiuraW} the variation of the global
charges readily reduces to 
\begin{equation}
\delta{\cal Q}^{\pm}=\mp\int d\varphi\left(\varepsilon_{\pm}\delta{\cal L}_{\pm}-\chi_{\pm}\delta{\cal W}_{\pm}\right)=\mp\frac{k}{4\pi}\int d\varphi\left({\cal \eta}^{\pm}\delta{\cal J}^{\pm}+\frac{4}{3}{\cal \eta}_{\left(3\right)}^{\pm}\delta{\cal J}_{\left(3\right)}^{\pm}\right)\;,
\end{equation}
so that they satisfy the current algebras in \eqref{algebra3}. Indeed,
this result just reflects the fact that the gauge transformation that
maps our asymptotic conditions in the highest weight and diagonal
gauges, spanned by the group element defined through \eqref{g1mn-g2mn},
is a permissible one in the sense of \cite{Bunster:2014mua}. Therefore, the
global charges associated to our asymptotic conditions, although written
in the highest weight gauge as in eqs. \eqref{eq:MuJ}, \eqref{eq:MuU}
and \eqref{eq:MiuraL}, \eqref{eq:MiuraW}, manifestly do not fulfill
the $W_{3}$ algebra. This is because the Lagrange multipliers $\mu_{\pm}$,
$\nu_{\pm}$, are not chosen to be fixed at infinity without variation
as in \cite{Henneaux:2013dra,Bunster:2014mua}, but instead, here they explicitly depend
on the global charges. What is actually kept fixed at the boundary
without variation is the set of Lagrange multipliers that is naturally
defined in the diagonal gauge ($\delta\zeta^{\pm}=\delta\zeta_{\left(3\right)}^{\pm}=0$).

\section{Higher spin black holes with soft hair}
\label{higher spin black hole}
In this section we discuss higher spin black hole solutions with soft hair, address their regularity and calculate their entropy. In section \ref{sec:regul-black-hole} we present our results in diagonal gauge, which is most suitable for our boundary conditions. In section \ref{sec:remarks-regul-entr} we discuss our results in highest weight gauge that is used traditionally in higher spin theories.

\subsection{Regularity and black hole entropy in diagonal gauge}
\label{sec:regul-black-hole}

As shown in section \ref{soft hair}, the simpler subset of our boundary
conditions, obtained by choosing the Lagrange multipliers $\zeta^{\pm}$,
$\zeta_{\left(3\right)}^{\pm}$ to be constants, possesses the noticeable
property of making the global charges $J_{n}^{\pm}$, $J_{m}^{\left(3\right)\pm}$
to behave as (higher spin) soft hair. An additional
remarkable feature that also occurs in this case is the fact
that regularity of the whole spectrum of Euclidean solutions that
fulfill our boundary conditions holds everywhere, regardless the value
of the global charges. 

The entire space of solutions of the field equations \eqref{FE-DG}
that satisfies our boundary conditions in this case is given by ${\cal J}^{\pm}={\cal J}^{\pm}\left(\varphi\right)$,
${\cal J}_{\left(3\right)}^{\pm}={\cal J}_{\left(3\right)}^{\pm}\left(\varphi\right)$,
which generically describes stationary non spherically symmetric higher
spin black flowers endowed with all the possible
left and right $\hat{u}\left(1\right)$ charges. In order to see this
explicitly, we assume that the topology of the Euclidean manifold
is the one of a solid torus, where the Euclidean time coordinate $\tau=it$
corresponds to the contractible cycle. As explained in \cite{Henneaux:2013dra,Bunster:2014mua},
the Lagrange multipliers correspond to the chemical potentials associated
to the global charges (see below), and since in our case they are
all switched on, the range of the coordinates can be fixed once and
for all. Here we assume that the boundary of the solid torus is described
by a trivial modular parameter, so that $0\leq\varphi<2\pi$, $0\leq\tau<\beta$,
where $\beta=T^{-1}$ is the inverse of the Hawking temperature.

Regularity of the Euclidean solution then requires the holonomy of
the gauge fields along any contractible cycle ${\cal C}$ to be trivial,
which reads
\begin{equation}
{\cal H}_{{\cal C}}={\cal P}e^{\int_{{\cal C}}a}=\mathbbm{1}\;.\label{HoloConds}
\end{equation}
Since the general solution is naturally formulated in the diagonal
gauge, as in \eqref{amn}, the regularity condition \eqref{HoloConds}
is trivially solved by
\begin{equation}
\zeta^{\pm}=\frac{2\pi n}{\beta}+\frac{2}{3}{\cal \zeta}_{\left(3\right)}^{\pm} \qquad \qquad {\cal \zeta}_{\left(3\right)}^{\pm}=\frac{3}{2}\frac{\pi}{\beta}m\;,\label{zetamn}
\end{equation}
with $m$, $n$ being integers. Hence, regularity
holds regardless the value of the global charges.

The generic regular solution then carries the entire set
of $\hat{u}\left(1\right)$ charges spanned by $J_{n}^{\pm}$ , $J_{m}^{\left(3\right)\pm}$.
Therefore, by virtue of the map between the $\hat{u}\left(1\right)$
and $W_{3}$ currents, defined through \eqref{eq:MiuraL}, \eqref{eq:MiuraW},
this class of solutions is endowed with nontrivial spin-2 and spin-3
(composite) charges.

This class of static non-spherically symmetric solutions are higher spin black flowers, whose ripples cannot
be gauged away because they are characterized by their corresponding
(higher spin) soft hair charges.

The simplest case in which the solution is only endowed with
the zero-modes charges $J_{0}^{\pm}$ , $J_{0}^{\left(3\right)\pm}$,
reduces to the stationary spherically symmetric higher spin black
hole solution discussed in \cite{Henneaux:2013dra,Bunster:2014mua}.

Generic higher spin black holes with
soft hair can be obtained from the spherically symmetric one aforementioned
by acting on it with an arbitrary soft boost, which corresponds
to applying a generic element of the asymptotic symmetry group globally.
Thus, as explained in section \ref{soft hair}, the action of the
$\hat{u}\left(1\right)$ generators does not change the total energy,
and noteworthy, the generic solution obtained through this procedure
remains regular. This is a peculiar feature of our boundary conditions,
which does not hold for different choices of Lagrange multipliers.
Indeed, for the choice in \cite{Henneaux:2013dra,Bunster:2014mua}, in which the Lagrange
multipliers in the highest weight are chosen to be constant, if one
applies the same procedure starting from a spherically symmetric solution,
one finds that under the action of the $W_{3}$ symmetries not only
the energy changes, but the regularity of the $W_{3}$-boosted
solution is generically spoiled. This reflects the fact that the additional
$W_{3}$ charges of the boosted solution do not correspond to soft
hair, since they do not commute with the Hamiltonian.

As shown in \cite{Perez:2012cf,Perez:2013xi}, the correct expression
for the black hole entropy in the context of higher spin gravity can
be obtained from 
\begin{align}
\delta S & =\beta\delta H=-\frac{k_{N}}{\pi}\text{Im}\left(\beta\int d\varphi\text{tr}\left[a_{\tau}\delta a_{\varphi}\right]_{\text{ }}\right)\;.\label{deltaS}
\end{align}
Under certain reasonable assumptions, which are fulfilled by our set
of boundary conditions expressed in the diagonal gauge, according
to \cite{deBoer:2013gz,Bunster:2014mua}, the variation of the entropy integrates
as 
\begin{align}
S & =-\frac{k_{N}}{\pi}\text{Im}\left(\beta\int d\varphi\text{tr}\left[a_{\tau}a_{\varphi}\right]_{\text{ }}\right)\;.\label{S}
\end{align}
The entropy of a generic higher spin black flower can then be readily
obtained from \eqref{S}, which by virtue of \eqref{amn} and
\eqref{zetamn}, in the Lorentzian case reads
\begin{equation}
S=\pi\left[\left(2n+m\right)\left(J_{0}^{+}+J_{0}^{-}\right)+3m\left(J_{0}^{\left(3\right)+}+J_{0}^{\left(3\right)-}\right)\right]\;.\label{Smn}
\end{equation}
Note that the entropy in \eqref{Smn} only captures the electric-like
zero-mode charges, and hence it does not depend on (higher spin) soft
hair.

An interesting effect occurs for the
branch of higher spin black flowers that is continuously connected
to the BTZ black hole \cite{Banados:1992wn,Banados:1992gq}, corresponding to $m=0$, $n=1$. Indeed, for
this branch the entropy \eqref{Smn} is found to depend just on the
zero modes of the electric-like $\hat{u}\left(1\right)$ charges of
the purely gravitational sector, i.e.,
\begin{equation}
S=2\pi\left(J_{0}^{+}+J_{0}^{-}\right)\;.\label{SbtzBranch}
\end{equation}
Nonetheless, the information about the presence of the higher spin
fields is subtlety hidden within the purely gravitational global charges,
as it can be seen from the map between the $\hat{u}\left(1\right)$
and $W_{3}$ currents. In fact, for the spherically symmetric higher
spin black hole, by virtue of \eqref{eq:MiuraL}, \eqref{eq:MiuraW},
the relationship between the zero modes of the purely gravitational
$\hat{u}\left(1\right)$ charges and the zero modes of the $W_{3}$
ones reads
\begin{equation}
J_{0}^{\pm}=\sqrt{2\pi k\mathcal{L}_{\pm}}\cos\left[\frac{1}{3}\arcsin\left(\frac{3}{8}\sqrt{\frac{3k}{2\pi\mathcal{L}_{\pm}^{3}}}\mathcal{W}_{\pm}\right)\right]\;.\label{Jo-LW}
\end{equation}
Therefore, replacing \eqref{Jo-LW} into \eqref{SbtzBranch} one recovers
the following expression for the higher spin black hole entropy in
terms of the spin-2 and spin-3 charges, which reads
\begin{align}
S & =2\pi\sqrt{2\pi k}\left(\sqrt{\mathcal{L}_{+}}\cos\left[\frac{1}{3}\arcsin\left(\frac{3}{8}\sqrt{\frac{3k}{2\pi\mathcal{L}_{+}^{3}}}\mathcal{W}_{+}\right)\right]\right.\nonumber \\
 & \quad \left.+\sqrt{\mathcal{L}_{-}}\cos\left[\frac{1}{3}\arcsin\left(\frac{3}{8}\sqrt{\frac{3k}{2\pi\mathcal{L}_{-}^{3}}}\mathcal{W}_{-}\right)\right]\right)\;,
\end{align}
in full agreement with the result obtained in \cite{Bunster:2014mua}.

\subsection{Remarks on regularity and entropy in highest weight gauge}
\label{sec:remarks-regul-entr}

As explained in section \ref{Mapping}, our boundary conditions can
be expressed in terms of the natural variables of the highest weight
gauge by choosing the Lagrange multipliers $\mu_{\pm}$, $\nu_{\pm}$
according to \eqref{eq:MuJ}, \eqref{eq:MuU}, as well as mapping
the $W_{3}$ currents ${\cal L}_{\pm}$, ${\cal W}_{\pm}$ in terms
of the $\hat{u}\left(1\right)$ ones as in \eqref{eq:MiuraL}, \eqref{eq:MiuraW}.
Despite that the formulation of our boundary conditions in the highest
weight gauge is certainly more involved than in the diagonal one,
it is worth mentioning that the computations related to regularity
of the Euclidean higher spin black flowers and their entropy can be
carried out anyway, by taking into account certain refinements. Indeed,
in the highest weight gauge, the left hand side of the regularity
condition \eqref{HoloConds} cannot be so easily exponentiated. Nevertheless,
the regularity condition can be alternatively written as 
\begin{align}
\beta^{2}\text{tr}\left[\hat{a}_{\tau}^{2}\right]&=  -8\pi^{2}\left(m^{2}+mn+n^{2}\right)\\
\beta^{3}\text{det}\left[\hat{a}_{\tau}\right]&= 8\pi^{3}i\left(m+n\right)mn\;,\label{Holo-HW}
\end{align}
where $\hat{a}_{\tau}$ stands for the Euclidean continuation of $\hat{a}_{t}^{\pm}$
in \eqref{amnHW}, with $\Lambda^{\pm}\left[\mu_{\pm},\nu_{\pm}\right]$
given by \eqref{Lambda-HW}. The latter equations actually become
extremely convoluted, since $\text{tr}\left[\hat{a}_{\tau}^{2}\right]$
and $\text{det}\left[\hat{a}_{\tau}\right]$ evaluate as 
\begin{align}
\text{tr}\left[\hat{a}_{\tau}^{2}\right]= & \frac{48\pi}{k}\left(\nu\mathcal{W}-\frac{1}{3}\mu\mathcal{L}+\frac{k}{12\pi}\mu^{\prime\prime}\right)\mu-\frac{2^{9}\pi^{2}}{3k^{2}}\left[\left(\mathcal{L}^{2}-\frac{k}{16\pi}\mathcal{L}^{\prime\prime}\right)\nu\right.\\
 & \left.-\frac{5k}{2^{5}\pi}\left(\nu^{\prime}\mathcal{L}^{\prime}+2\nu^{\prime\prime}\mathcal{L}-\frac{k}{20\pi}\nu^{\prime\prime\prime\prime}\right)\right]\nu-\frac{2}{3}\nu^{\prime\prime2}-2\mu^{\prime2}+\frac{4}{3}\left(\nu^{\prime\prime\prime}-\frac{20\pi}{k}\nu^{\prime}\mathcal{L}\right)\nu^{\prime} \, , \nonumber
\end{align}
\begin{align}
i\text{det}\left[\hat{a}_{\tau}\right] & =\frac{2^{7}\pi^{2}}{3k^{2}}\left[\nu^{2}\mathcal{L}-\frac{3k}{32\pi}\left(\mu^{2}-\nu^{\prime2}+\frac{4}{3}\nu\nu^{\prime\prime}\right)\right]\left[\mu\mathcal{W}+\frac{2}{3}\nu\left(\mathcal{L}^{\prime\prime}-\frac{12\pi}{k}\mathcal{L}^{2}\right)\right. \nonumber\\
 & \quad \left.+\frac{7}{3}\nu^{\prime}\mathcal{L}^{\prime}+\frac{8}{3}\nu^{\prime\prime}\mathcal{L}-\frac{k}{12\pi}\nu^{\prime\prime\prime\prime}\right]+\frac{2^{5}\pi^{2}}{3k^{2}}\left[\left(\mu-\frac{5}{3}\nu^{\prime}\right)\mathcal{L}-2\left(\mathcal{W}+\frac{1}{3}\mathcal{L}^{\prime}\right)\nu\right. \nonumber\\
 & \quad \left.-\frac{k}{4\pi}\left(\mu-\frac{1}{3}\nu^{\prime}\right)^{\prime\prime}\right]\left[\left(\mu-\nu^{\prime}\right)\left(\nu\mathcal{L}-\frac{3k}{8\pi}\left(\mu+\frac{1}{3}\nu^{\prime}\right)^{\prime}\right)\right.\\
 & \quad \left.-6\nu\left(\left(\mathcal{W}-\frac{1}{3}\mathcal{L}^{\prime}\right)\nu-\frac{1}{2}\left(\mu+\frac{5}{3}\nu^{\prime}\right)\mathcal{L}+\frac{k}{8\pi}\left(\mu+\frac{1}{3}\nu^{\prime}\right)^{\prime\prime}\right)\right] \nonumber\\
 & \quad +\frac{2^{3}\pi}{k}\left(\mu^{\prime}-\frac{1}{3}\left(\nu^{\prime\prime}-\frac{8\pi}{k}\nu\mathcal{L}\right)\right)\left[\left(\mu+\nu^{\prime}\right)\left(\nu\left(\mathcal{W}-\frac{1}{3}\mathcal{L}^{\prime}\right)-\frac{1}{2}\left(\mu+\frac{5}{3}\nu^{\prime}\right)\mathcal{L}\right.\right.\nonumber\\
 & \quad \left.\left.+\frac{k}{8\pi}\left(\mu+\frac{1}{3}\nu^{\prime}\right)^{\prime\prime}\right)-\frac{2^{4}\pi}{3^{2}k}\left(\nu\mathcal{L}-\frac{k}{8\pi}\nu^{\prime\prime}\right)\left(\nu\mathcal{L}-\frac{3k}{8\pi}\left(\mu+\frac{1}{3}\nu^{\prime}\right)^{\prime}\right)\right]\;, \nonumber
\end{align}
respectively. Nonetheless, after some amount of algebraic work, one
verifies that the regularity conditions in \eqref{Holo-HW} 
conspire with the expressions that define our boundary conditions
in the highest weight gauge, given by \eqref{eq:MuJ}, \eqref{eq:MuU},
and \eqref{eq:MiuraL}, \eqref{eq:MiuraW}, so that regularity holds
 provided the equations in \eqref{zetamn} do, in full agreement with
the result that was easily performed in the diagonal gauge.

The formulation of our boundary
conditions in the highest weight gauge is such that the hypotheses
assumed in \cite{deBoer:2013gz,Bunster:2014mua} do not apply, and hence the
entropy formula in \eqref{S} cannot be used in this case, since it
would yield an incorrect result. However, the higher spin black flower
entropy can still be obtained through the original formula in \eqref{deltaS}
\cite{Perez:2012cf,Perez:2013xi}. Indeed, replacing $\hat{a}_{\varphi}^{\pm}$
and $\hat{a}_{t}^{\pm}$ into the variation of the (Lorentzian) entropy,
one obtains 
\begin{align*}
\delta S & =\frac{k_{3}}{2\pi}\beta\int d\varphi\left(\text{tr}\left[\hat{a}_{\tau}^{+}\delta\hat{a}_{\varphi}^{+}\right]_{\text{ }}-\text{tr}\left[\hat{a}_{\tau}^{-}\delta\hat{a}_{\varphi}^{-}\right]_{\text{ }}\right) \\
 & =\int d\varphi\left(\mu_{+}\delta{\cal L}_{+}+\mu_{-}\delta{\cal L}_{-}-\nu_{+}\delta{\cal W}_{+}-\nu_{-}\delta{\cal W}_{-}\right)\;,
\end{align*}
which by virtue of \eqref{eq:MuJ}, \eqref{eq:MuU}, and \eqref{eq:MiuraL},
\eqref{eq:MiuraW}, reduces to 
\begin{equation}
\delta S=\frac{k_{3}}{4\pi}\beta\int d\varphi\left(\zeta^{+}\delta\mathcal{J}^{+}+\zeta^{-}\delta\mathcal{J}^{-}+\frac{4}{3}\zeta_{\left(3\right)}^{+}\delta\mathcal{J}_{\left(3\right)}^{+}+\frac{4}{3}\zeta_{\left(3\right)}^{-}\delta\mathcal{J}_{\left(3\right)}^{-}\right)\;.
\end{equation}
Since, $\zeta^{\pm}$ and $\zeta_{\left(3\right)}^{\pm}$ are assumed
to be constants, the variation of the entropy readily integrates as
\begin{equation}
S=\frac{k_{3}}{4\pi}\beta\int d\varphi\left(\zeta^{+}\mathcal{J}^{+}+\zeta^{-}\mathcal{J}^{-}+\frac{4}{3}\zeta_{\left(3\right)}^{+}\mathcal{J}_{\left(3\right)}^{+}+\frac{4}{3}\zeta_{\left(3\right)}^{-}\mathcal{J}_{\left(3\right)}^{-}\right)\;,
\end{equation}
so that once the regularity conditions in \eqref{zetamn} are taken
into account, the entropy reduces to the formula in eq. \eqref{Smn}
which was straightforwardly obtained in the diagonal gauge.

\section{Higher spin black holes with soft hair in the metric formalism}
\label{sec:higher-spin-black}

The entropy of the higher spin black flower for the branch that is
connected with the BTZ black hole in eq. \eqref{SbtzBranch} can
also be alternatively recovered in the metric formalism. In order
to deal with non spherically symmetric horizons, the formula proposed
in \cite{Perez:2013xi} has to be slightly refined, so that the entropy
reads 
\begin{equation}
S=\frac{1}{4G}\int_{\partial\Sigma_{+}}{\cal A}\cos\left[\frac{1}{3}\arcsin\left(\left(\sqrt{3}\frac{\phi}{{\cal A}}\right)^{3}\right)\right]d\sigma\ ,\label{S-Metric}
\end{equation}
where ${\cal A}$ and $\phi$ stand for the horizon area element and
its spin-3 analogue, respectively. They are naturally defined in terms
of the pullback of the metric and the spin-3 field at the spacelike
section of the horizon, so that their integrals become reparametrization
invariant, i.e.,
\begin{align}
A&=\int_{\partial\Sigma_{+}}{\cal A} \, d\sigma=\int_{\partial\Sigma_{+}}\left(g_{\mu\nu}\frac{dx^{\mu}}{d\sigma}\frac{dx^{\nu}}{d\sigma}\right)^{1/2}d\sigma \label{Area}
\\
\Phi&=\int_{\partial\Sigma_{+}}\phi \, d\sigma=\int_{\partial\Sigma_{+}}\left(\Phi_{\mu\nu\rho}\frac{dx^{\mu}}{d\sigma}\frac{dx^{\nu}}{d\sigma}\frac{dx^{\rho}}{d\sigma}\right)^{1/3}d\sigma\ .\label{spin-3-Area}
\end{align}
It is worth mentioning that, although a full nonperturbative action
principle for higher spin gravity formulated exclusively in term of
the metric and the spin-3 field is still unknown, the entropy in \eqref{S-Metric}
can be written as a closed analytic formula. In the weak
spin-3 limit, an explicit action principle was constructed in \cite{Campoleoni:2012hp}
up to quadratic order, from which the entropy in the static case was
also obtained and it turns out to agree with the corresponding perturbative
expansion of \eqref{S-Metric}.

The higher spin black flower metric corresponds to a generalization
of the soft hairy black holes recently obtained in \cite{Afshar:2016wfy}
for General Relativity with negative cosmological constant in vacuum.
In order to explicitly see the contact, for simplicity, let us choose
the Lagrange multipliers to be constants according to $\zeta^{\pm}=-a$,
${\cal \zeta}_{\left(3\right)}^{\pm}=-a_{\left(3\right)}$. The spacetime
metric can then be reconstructed from the gauge fields as in \eqref{MetricAndSpin-S-Fields},
and it is found to be given by 
\begin{align}
ds^{2}= & -2\rho\ell\mathit{f}\left(\rho\right)\mathit{a}\left(1+4\left(1-2\mathit{f}\left(\rho\right)\right)^{2}a^{-2}a_{\left(3\right)}^{2}\right)dv^{2}+2\ell d\rho dv-2\omega a^{-1}d\rho d\varphi\nonumber \\
 & +4\left(\omega+4\left(1-2\mathit{f}\left(\rho\right)\right)^{2}\omega_{\left(3\right)}a^{-1}a_{\left(3\right)}\right)\rho\mathit{f}\left(\rho\right)dvd\varphi\label{Metric-Null}\\
 & +\left[\gamma^{2}+\frac{4}{3}\gamma_{\left(3\right)}^{2}+\frac{2\rho}{\ell}a^{-1}\mathit{f}\left(\rho\right)\left(\gamma^{2}-\omega^{2}+4\left(1-2\mathit{f}\left(\rho\right)\right)^{2}\left(\gamma_{\left(3\right)}^{2}-\omega_{\left(3\right)}^{2}\right)\right)\right]d\varphi^{2}\;,\nonumber 
\end{align}
with $t=v$, and $f\left(\rho\right)=1+\frac{\rho}{2a\ell}$. The
remaining arbitrary functions are related to the global charges according
to
\begin{equation}
\ell{\cal J}^{\pm}:=\gamma\pm\omega\;\;;\;\;\ell{\cal J}_{\left(3\right)}^{\pm}:=\gamma_{\left(3\right)}\pm\omega_{\left(3\right)}\;,
\end{equation}
which means that they cannot be gauged away by proper gauge transformations.
The event horizon is located at $\rho=0$. The line-element \eqref{Metric-Null} shows that our boundary conditions can be interpreted as near horizon conditions that for small $\rho$ recover Rindler space in Eddington-Finkelstein type of coordinates.
In the case of $\omega_{\left(3\right)}=a_{\left(3\right)}=0$, the
spacetime geometry reduces to the one found in \cite{Afshar:2016wfy}
which is of negative constant curvature. 

Analogously, the spin-3 field can be reconstructed from \eqref{MetricAndSpin-S-Fields},
whose explicit form is given in the Appendix \ref{Appendix:B}. For
our purposes, it is enough to know the behaviour of its purely angular
component around the horizon, which reads
\begin{equation}
\Phi_{\varphi\varphi\varphi}=-\frac{2}{27}\left(4\gamma_{\left(3\right)}^{2}-9\gamma^{2}\right)\gamma_{\left(3\right)}+{\cal O}\left(\rho\right)\ .\label{phi-Null}
\end{equation}

Note that in three spacetime dimensions the metric component $g_{\rho\varphi}$
can always be gauged away. Indeed, even in the case of a generic choice
of Lagrange multipliers $\zeta^{\pm}$, $\zeta_{\left(3\right)}^{\pm}$,
the metric and the spin-3 field can also be directly reconstructed
from \eqref{MetricAndSpin-S-Fields} in normal coordinates. This can
be done through a permissible gauge transformation, which amounts
to replace the group elements $b_{\pm}$ in \eqref{b-null} by $b_{\pm}=e^{\pm\frac{r}{2}\left(\Lt_{1}+\Lt_{-1}\right)}$.
The line element then reads
\begin{align}
ds^{2}= & -\frac{\ell^{2}}{2}\left[\zeta_{\left(3\right)}^{+}\zeta_{\left(3\right)}^{-}\left(\cosh\left(4r\right)+\frac{5}{3}\right)+2\zeta^{+}\zeta^{-}\cosh^{2}\left(r\right)-\frac{1}{2}\left(\zeta^{+}+\zeta^{-}\right)^{2}\right.\nonumber \\
 & \left.-\frac{2}{3}\left(\zeta_{\left(3\right)}^{+}+\zeta_{\left(3\right)}^{-}\right)^{2}\right]dt^{2}+\ell^{2}dr^{2}+\ell\left[\frac{1}{2}\left(\zeta_{\left(3\right)}^{+}-\zeta_{\left(3\right)}^{-}\right)\gamma_{\left(3\right)}\left(\cosh\left(4r\right)+\frac{5}{3}\right)\right.\nonumber \\
 & \left.-\left(\zeta_{\left(3\right)}^{+}+\zeta_{\left(3\right)}^{-}\right)\omega_{\left(3\right)}\text{sinh}^{2}\left(2r\right)-\left(\zeta^{+}+\zeta^{-}\right)\omega\text{sinh}^{2}\left(r\right)+\left(\zeta^{+}-\zeta^{-}\right)\gamma\text{cosh}^{2}\left(r\right)\right]dtd\varphi\nonumber \\
 & +\left[\frac{1}{2}\left(\cosh\left(4r\right)+\frac{5}{3}\right)\gamma_{\left(3\right)}^{2}-\omega_{\left(3\right)}^{2}\text{sinh}^{2}\left(2r\right)+\gamma^{2}\cosh^{2}\left(r\right)-\omega^{2}\text{sinh}^{2}\left(r\right)\right]d\varphi^{2}\;,\label{Metric-NormalC}
\end{align}
so that the event horizon locates at $r=0$. This class of geometries
asymptotically approaches to locally AdS$_{3}$ spacetimes of radius
$\ell/2$.

The explicit form of the spin-3 field is written in the Appendix \ref{Appendix:B},
and its purely angular component close to the horizon behaves as equation \eqref{phi-Null} with ${\cal O}(\rho)$ replaced by ${\cal O}(r^2)$.
It is then simple to verify that requiring regularity of the Euclidean
metric and the spin-3 field around the horizon fixes the Lagrange
multipliers (chemical potentials) precisely as in eq. \eqref{zetamn}
with $m=0$ and $n=1$, i.e., for the branch that is connected to
the BTZ black hole. Further issues about regularity of the fields
in the metric formalism have been discussed in \cite{Gutperle:2011kf,Banados:2016nkb}.

For the most generic higher spin black flower configuration \eqref{Metric-NormalC}, and (the again transformed) \eqref{phi-Null},
as well as for the particular case in \eqref{Metric-Null},
\eqref{phi-Null}, the event horizon area element and its spin-3 analogue
in eqs. \eqref{Area}, \eqref{spin-3-Area}, respectively, then evaluate
as
\begin{align}
{\cal A}^{2} & =\left.g_{\varphi\varphi}\right\vert _{r_{+}}=\gamma^{2}+\frac{4}{3}\gamma_{\left(3\right)}^{2} \label{Area-blackflower}\\
\phi^{3} & =\left.\Phi_{\varphi\varphi\varphi}\right\vert _{r_{+}}=\frac{2}{27}\gamma_{\left(3\right)}\left(9\gamma^{2}-4\gamma_{\left(3\right)}^{2}\right)\;,\label{spin-3-area-blackflower}
\end{align}
so that, by virtue of the identity
\begin{equation}
\cos\left[\frac{1}{3}\arcsin\left(\frac{x\left(x^{2}-3\right)}{\left(x^{2}+1\right)^{3/2}}\right)\right]=\frac{1}{\sqrt{x^{2}+1}} 
\end{equation}
the entropy in \eqref{S-Metric} reduces to
\begin{align}
S & =\frac{1}{4G}\int\gamma \, d\varphi=\frac{k}{2}\int\left({\cal J}^{+}+{\cal J}^{-}\right)d\varphi=2\pi\left(J_{0}^{+}+J_{0}^{-}\right)\;,
\end{align}
in full agreement with the result found exclusively in terms of gauge
fields in section \ref{higher spin black hole} (see eq. \eqref{SbtzBranch}).

Note that $\int\gamma d\varphi$ could then be regarded as a sort
of ``effective higher spin horizon area'', which curiously corresponds
to the horizon area of the purely gravitational black flower in the
absence of higher spin fields.

\section{Extension to AdS$_{3}$ gravity coupled to fields of spin greater
than 2}
\label{GenericN}

The extension of our boundary conditions and the analysis of the properties
of the corresponding higher spin black flower can also be performed
in the case of gravity with negative cosmological constant coupled
to bosonic fields of spin $s > 2$. As explained in the introduction,
the generic theory is described by a gauge group given by two copies
of $sl\left(N,\mathbb{R}\right)$. It is also worth mentioning that
in the case of even $N$, the theory admits an additional truncation
that describes the coupling of gravitation on AdS$_{3}$ with fields
of even spin $s=4$, $6$, $\cdots$, $N$, described by two copies
of $sp\left(N,\mathbb{R}\right)$. Additional special truncations
are also known to exist in the case of exceptional groups, see e.g.,
\cite{Chen:2012pc}. Hereafter we
carry out the analysis for $sl\left(N,\mathbb{R}\right)$.

We propose that the asymptotic behaviour of the $sl\left(N,\mathbb{R}\right)$
gauge fields $A^{\pm}$ is of the form in \eqref{Amn} with $b_{\pm}$
given by \eqref{b-null}, so that the auxiliary connection extends
to
\begin{equation}
a^{\pm}=\left(\pm\mathcal{J}^{\pm}\ d\varphi+\zeta^{\pm}\ dt\right)\Lt_{0}+\sum_{s=3}^{N}\left(\pm\mathcal{J}_{\left(s\right)}^{\pm}\ d\varphi+\zeta_{\left(s\right)}^{\pm}\ dt\right)\Wt_{0}^{\left(s\right)}\;,\label{eq:achicoN}
\end{equation}
where $\Lt_{0}$, and $\Wt_{0}^{\left(s\right)}$ stand for the generators
of the Cartan subalgebra of the gauge group. The dynamical fields
are then described by arbitrary functions of time and the angular
coordinate, given by $\mathcal{J}^{\pm}$, $\mathcal{J}_{\left(s\right)}^{\pm}$,
with $s=3$, $4$, $\cdots$, $N$; and their corresponding Lagrange
multipliers, $\zeta^{\pm}$, $\zeta_{\left(s\right)}^{\pm}$, are
assumed to be fixed at the boundary without variation.
It is then simple to verify that the field equations read 
\begin{equation}
\dot{\mathcal{J}}^{\pm}=\pm\zeta^{\prime} \qquad \qquad \dot{\mathcal{J}}_{\left(s\right)}^{\pm}=\pm\zeta_{\left(s\right)}^{\prime}\;.\label{FE-DG-1}
\end{equation}

The asymptotic symmetries that maintain the form of \eqref{eq:achicoN}
are spanned by Lie-algebra-valued parameters of the form $\epsilon^{\pm}=\eta^{\pm}\Lt_{0}+\sum_{s=3}^{N}\eta_{\left(s\right)}^{\pm}\Wt_{0}^{\left(s\right)}$,
with $\dot{\eta}^{\pm}=\dot{\eta}_{\left(s\right)}^{\pm}=0$, provided
the fields transform according to 
\begin{equation}
\delta\mathcal{J}^{\pm}=\pm\eta^{\pm\prime}\qquad \qquad\delta\mathcal{J}_{\left(s\right)}^{\pm}=\pm\eta_{\left(s\right)}^{\pm\prime}\;.\label{TransfLaw JmnJ3mn-1-1}
\end{equation}
The canonical generators of the asymptotic symmetries then read
\begin{equation}
{\cal Q}^{\pm}\left[\eta^{\pm},\eta_{\left(s\right)}^{\pm}\right]=\mp\frac{k}{4\pi}\int d\varphi\left(\eta^{\pm}\left(\varphi\right)\mathcal{J}^{\pm}\left(\varphi\right)+\sum_{s=3}^{N}\alpha_{s}\eta_{\left(s\right)}^{\pm}\left(\varphi\right)\mathcal{J}_{\left(s\right)}^{\pm}\left(\varphi\right)\right),
\end{equation}
where
\begin{equation}
\alpha_{s}=\frac{48\left(s-1\right)!^{4}}{\left(2s-1\right)!\left(2s-2\right)!}\prod_{i=2}^{s-1}\left(N^{2}-i^{2}\right)
\end{equation}
and their algebra is given by
\begin{align}
\left\{ \mathcal{J}^{\pm}\left(\varphi\right),\mathcal{J}^{\pm}\left(\varphi^{\prime}\right)\right\}  & =\mp\left(\frac{4\pi}{k}\right)\delta^{\prime}\left(\varphi-\varphi^{\prime}\right) \nonumber \\
\left\{ \mathcal{J}_{\left(s\right)}^{\pm}\left(\varphi\right),\mathcal{J}_{\left(s^{\prime}\right)}^{\pm}\left(\varphi^{\prime}\right)\right\}  & =\mp\left(\frac{4\pi}{\alpha_{s}k}\right)\delta_{s,s^{\prime}}\delta^{\prime}\left(\varphi-\varphi^{\prime}\right)\;.\label{eq:algebraspins}
\end{align}
Expanding in Fourier modes according to
\begin{equation}
\mathcal{J}^{\pm}\left(\varphi\right)=\frac{2}{k}\sum_{n=-\infty}^{\infty}J_{n}^{\pm}e^{\pm in\varphi} \qquad \qquad \mathcal{J}_{\left(s\right)}^{\pm}\left(\varphi\right)=\frac{2}{\alpha_{s}k}\sum_{n=-\infty}^{\infty}J_{n}^{\left(s\right)\pm}e^{\pm in\varphi}\;,
\end{equation}
the algebra \eqref{eq:algebraspins} reads
\begin{equation}
\label{eq:spinNalg}
i\left\{ J_{n}^{\pm},J_{m}^{\pm}\right\} =\frac{k}{2}n\delta_{m+n,0} \qquad \qquad i\left\{ J_{n}^{\left(s\right)\pm},J_{m}^{\left(s^{\prime}\right)\pm}\right\} =\frac{\alpha_{s}k}{2}n\delta_{s,s^{\prime}}\delta_{m+n,0}\;,
\end{equation}
which corresponds to a set of $\hat{u}\left(1\right)$ currents with
levels $\frac{1}{2}k$ and $\frac{1}{2}\alpha_{s}k$.

For the simplest choice of boundary conditions, given by constant
Lagrange multipliers $\zeta^{\pm}$, $\zeta_{\left(s\right)}^{\pm}$,
the dynamical fields become time-independent, and the total Hamiltonian
reads
\begin{equation}
H=\zeta^{+}J_{0}^{+}+\zeta^{-}J_{0}^{-}+\sum_{s=3}^{N}\left(\zeta_{\left(s\right)}^{+}J_{0}^{\left(s\right)+}+\zeta_{\left(s\right)}^{-}J_{0}^{\left(s\right)-}\right)\;,
\end{equation}
which manifestly commutes with the $\hat{u}\left(1\right)$ generators
$J_{n}^{\pm}$, $J_{m}^{\left(s\right)\pm}$. Therefore, in this case,
the whole set of affine global charges can be regarded as (higher
spin) soft hair in the sense of \cite{Hawking:2016msc}.

Following the procedure described in section \ref{Mapping}, one naturally
expects to find the generalization of the Miura transformation in
\eqref{eq:MiuraL}, \eqref{eq:MiuraW} between the currents in the
diagonal and highest weight gauges, from which the $W_{N}$ currents
should emerge from composite operators of the affine ones through
an analogue of the twisted Sugawara construction.

Note that in the case of constant $\zeta^{\pm}$ and $\zeta_{\left(s\right)}^{\pm}$, 
the field equations \eqref{FE-DG-1} imply that the space
of solutions that fulfills our boundary conditions is described by
${\cal J}^{\pm}={\cal J}^{\pm}\left(\varphi\right)$, ${\cal J}_{\left(s\right)}^{\pm}={\cal J}_{\left(s\right)}^{\pm}\left(\varphi\right)$,
which generically corresponds to non spherically symmetric higher
spin black holes endowed with all of the possible left and right soft
$\hat{u}\left(1\right)$ charges. Indeed, in the fundamental representation
of $sl\left(N,\mathbb{R}\right)$, regularity of the Euclidean configurations
reads 

\[
{\cal H}_{{\cal C}}={\cal P}e^{\int_{{\cal C}}a}=\left(-1\right)^{N+1}\mathbbm{1}\;,
\]
where ${\cal C}$ stands for any contractible cycle in the solid torus.
The regularity conditions of the Euclidean solutions then just reduces
to exponentiate a diagonal matrix, which readily implies that the
chemical potentials $\zeta^{\pm}$, $\zeta_{\left(s\right)}^{\pm}$
become fixed by precise relationships that depend linearly on $N-1$
arbitrary integers, and they are independent of the global charges.
For the branch that is connected to the BTZ black hole, the only nonvanishing
chemical potentials are the ones that correspond to the purely gravitational
sector, i.e., 
\begin{equation}
\zeta^{\pm}=\frac{2\pi}{\beta} \qquad \qquad {\cal \zeta}_{\left(s\right)}^{\pm}=0\;.
\end{equation}

The generic solution is then described by $N-1$ left and right soft
$\hat{u}\left(1\right)$ charges, from which the higher spin ones
can be reconstructed by virtue of the map between them and the $W_{N}$
currents.

For the branch that is continuously connected to the BTZ black hole,
the entropy of a generic higher spin black flower can be directly
recovered from \eqref{S}, which reduces to
\begin{equation}
S=2\pi\left(J_{0}^{+}+J_{0}^{-}\right)\;.\label{S-sl(N)}
\end{equation}
Note that the entropy \eqref{S-sl(N)} exactly agrees with the one
in \eqref{SbtzBranch} for the case of $sl\left(3,\mathbb{R}\right)$,
as well as for the expression found in \cite{Afshar:2016wfy} in the
case of gravity on AdS$_{3}$. In this sense, the entropy in \eqref{S-sl(N)}
is universal because it depends only on the zero modes of the
purely gravitational $\hat{u}\left(1\right)$ charges, regardless
the value of $N\geq2$.

\section{Discussion}
\label{sec:discussion}

The new set of boundary conditions \eqref{Amn}-\eqref{amn} (or more generally, \eqref{eq:achicoN}) for higher spin gravity on AdS$_3$ has a set of affine $\hat u(1)$ currents generating the asymptotic symmetry algebra with non-vanishing levels \eqref{algebra3} (or more generally \eqref{eq:spinNalg}).
The concept of spin emerges from the conformal
weight of the composite generators of the asymptotic $W$ algebra
that is recovered through a nonlinear analogue of the twisted Sugawara
construction \eqref{eq:MiuraL}, \eqref{eq:MiuraW}. 
Fourier decomposition of the $\hat u(1)$ charges leads to a tower of generators \eqref{eq:Jmodes} which can be interpreted as creation operators for negative integer indices. Acting with these creation operators on states, e.g. on the vacuum as in \eqref{eq:angelinajolie}, generates higher spin soft hair descendants that have the same energy as the original state, thereby generalizing corresponding spin-2 results~\cite{Hawking:2016msc} (see also~\cite{Banks:2014iha,Dvali:2015rea,Donnay:2015abr,Blau:2015nee,Averin:2016ybl,Kehagias:2016zry,Eling:2016xlx,Setare:2016jba,Setare:2016vhy,Averin:2016hhm}).


The generic solution of the field equations that fulfills our boundary
conditions describes stationary and non necessarily spherically symmetric
configurations, whose entropy \eqref{S-sl(N)} was shown to be independent of (higher
spin) soft hair. In the case of AdS$_{3}$ gravity coupled to spin-3
fields, these results were also explicitly recovered in the metric
formalism in section \ref{sec:higher-spin-black}. It would be interesting to explore further geometric aspects
of these higher spin black flowers along the lines of \cite{Barnich:2015dvt,Troessaert:2015syk},
as well as the possibility of performing a microstate counting for
their entropy as in \cite{Afshar:2016uax}.

We have generalized our results from spin-3 gravity in AdS$_3$ to general higher spin gravity in section \ref{GenericN}, but there is a number of issues that remain for further exploration. For instance, we did not explicitly provide the (twisted) Sugawara-like constructions for spins greater than 3. Moreover, we focused exclusively on the principal embedding of $sl(2,\mathbb{R})$ into $sl(N,\mathbb{R})$~\cite{Castro:2012bc,Afshar:2012hc}. Additionally, we did not consider Vasiliev-type of theories with infinite towers of spins, based on $hs(\lambda)$~\cite{Henneaux:2010xg,Gaberdiel:2011wb}. Finally, our focus was exclusively on AdS$_3$ backgrounds, but it is well-known that higher spin gravity allows also for non-AdS backgrounds~\cite{Gary:2012ms,Krishnan:2013zya,Gutperle:2013oxa,Gary:2014mca,Breunhoelder:2015waa} and vanishing cosmological constant~\cite{Afshar:2013vka,Gonzalez:2013oaa,Gary:2014ppa,Matulich:2014hea,prep16} or generalizations along the lines of hypergravity~\cite{Fuentealba:2015jma,Henneaux:2015ywa,Fuentealba:2015wza,Henneaux:2015tar}. 
Investigations of the other branches, besides the BTZ branch, of the entropy should be possible~\cite{David:2012iu}. Furthermore, it might be interesting to explore extremal higher spin black holes~\cite{Henneaux:2015ywa,Banados:2015tft} and entanglement entropy~\cite{Ammon:2013hba,deBoer:2013vca,Datta:2014ska,Castro:2014mza,deBoer:2014sna} in this setup.

Besides, a different set of boundary conditions for which the Lagrange
multipliers depend locally on the dynamical fields has been recently
proposed in \cite{Perez:2016vqo}, where it was shown that the field
equations correspond to different integrable hierarchies. Our set of boundary conditions, once expressed
in the highest weight gauge, differs from the one in \cite{Perez:2016vqo}
because here the Lagrange multipliers depend non-locally on the dynamical
fields (see e.g., eqs. \eqref{eq:MuJ}, \eqref{eq:MuU}, with \eqref{eq:MiuraL},
\eqref{eq:MiuraW}). Nonetheless, as pointed out in \cite{Perez:2016vqo},
for the case of pure gravity on AdS$_{3}$, our boundary conditions
appear to be related to a representative of an extension of the KdV
hierarchy that is labeled by a fractional instead of an integer label
(precisely $k=-1/2$), see e.g. \cite{FracKdV1,FracKdV2,FracKdV3,FracKdV4}.
One might then naturally expect that similar effect should occur for
the extension of different hierarchies once the higher spin fields
within our boundary conditions are switched on.

Additional interesting possibilities also naturally arise along the
lines of the connection of higher spin fields and string theory \cite{Gaberdiel:2014cha,Gaberdiel:2015wpo} and the diverse related results in refs. \cite{Tan:2011tj,Ferlaino:2013vga,Compere:2013gja,Compere:2013nba,Li:2013rsa,deBoer:2014fra,Cabo-Bizet:2014wpa,Cabo-Bizet:2014rpa,Li:2015osa,Apolo:2015zxh,Apolo:2016sui}.


\section*{Acknowledgments}
We thank H.~Afshar, G.~Barnich, X.~Bekaert, A.~Campoleoni, D.~Francia, O.~Fuentealba, M.~Gary,
J.~Matulich, W.~Merbis, M.~Riegler and J.~Simon for useful comments.
We also acknowledge the scientific atmosphere of the workshops
``Topics in three dimensional Gravity'' in March
2016 at the Abdus Salam ICTP, Trieste, Italy, the workshop ``Flat Holography'' in April 2016 at the Simons Center, Stony Brook, New York and the MIAPP programme ``Higher-Spin Theory and Duality'' in May 2016 at the Munich Institute for Astro- and Particle Physics, Germany.

The work of AP, DT and RT is partially funded by Fondecyt grants Nº
11130260, 11130262, 1130658, 1161311. The Centro de Estudios Científicos
(CECs) is funded by the Chilean Government through the Centers of
Excellence Base Financing Program of Conicyt. 
DG was supported by the Austrian Science Fund (FWF) projects P 27182-N27 and P 28751-N27 and by the program Science without Borders, project CNPq-401180/2014-0.
SP was supported by the FWF project P 27396-N27.

\appendix

\section{Principal embedding of $SL\left(2,\mathbb{R}\right)$ within $SL\left(N,\mathbb{R}\right)$}
\label{Appendix:A}

In the principal embedding of $SL\left(2,\mathbb{R}\right)$ within
$SL\left(N,\mathbb{R}\right)$, the set of generators of $sl\left(N,\mathbb{R}\right)$
naturally splits as $\{\Lt_{i};\Wt_{m}^{\left(s\right)}\}$, with $i=-1,0,1$;
$s=3,\cdots,N$, and $m=-\left(s-1\right),\cdots,\left(s-1\right)$,
so that
\begin{align}
\left[\Lt_{i},\Lt_{j}\right] & =\left(i-j\right)\Lt_{i+j} \\
\left[\Lt_{i},\Wt_{m}^{\left(s\right)}\right] & =\left(\left(s-1\right)i-m\right)\Wt_{i+m}^{\left(s\right)} 
\end{align}
from which $\Lt_{i}$ and $\Wt_{m}^{\left(s\right)}$ can be seen to
possess spin two and spin $s$, respectively. In the fundamental representation,
the generators of the $sl\left(2,\mathbb{R}\right)$ subset are described
through $N\times N$ matrices that can be chosen to be given by

\begin{align}
\left(\Lt_{1}\right)_{jk} & =-\sqrt{j\left(N-j\right)}\delta_{j+1,k} \\
\left(\Lt_{-1}\right)_{jk}&=\sqrt{k\left(N-k\right)}\delta_{j,k+1} \\
\left(\Lt_{0}\right)_{jk} & =\frac{1}{2}\left(N+1-2j\right)\delta_{j,k}\;,
\end{align}
with $j,k=2...,N$, so that the remaining ones become defined as
\begin{align}
\Wt_{m}^{\left(s\right)} & =2\left(-1\right)^{s-m-1}\frac{\left(s+m-1\right)!}{\left(2s-2\right)!}\underset{s-m-1\;\text{terms}}{\underbrace{[\Lt_{-1},[\Lt_{-1},\cdots[\Lt_{-1}}},\left(\Lt_{1}\right)^{s-1}]\cdots]] \\
 & =2\left(-1\right)^{s-m-1}\frac{\left(s+m-1\right)!}{\left(2s-2\right)!}\left(\mathrm{ad}_{\Lt_{-1}}\right)^{s-m-1}\left(\Lt_{1}\right)^{s-1}\;,
\end{align}
where $\mathrm{ad}_{\Xt}\left(\Yt\right):=\left[\Xt,\Yt\right]$.

The normalization constants used in the text, are then given by
\begin{align}
\epsilon_{N} & =\text{tr}(\Lt_{0}\Lt_{0})=\frac{N(N^{2}-1)}{12} \\
\alpha_{s} & =\frac{12}{N\left(N^{2}-1\right)}\text{tr}\left(\Wt_{0}^{\left(s\right)}\Wt_{0}^{\left(s\right)}\right)=\frac{48\left(s-1\right)!^{4}}{\left(2s-1\right)!\left(2s-2\right)!}\prod_{i=2}^{s-1}\left(N^{2}-i^{2}\right)\;.
\end{align}

\section{Explicit form of the spin-3 field}
\label{Appendix:B}

For the particular case of the solution that describes a higher spin
black flower, whose metric is given by \eqref{Metric-Null}, the explicit
form of the spin-3 field reads
\begin{align}
\Phi & =\left\{ a^{-2}\gamma_{\left(3\right)}d\rho^{2}+2\ell\left[\left(1-2\mathit{f}\left(\rho\right)\right)^{2}\left(a_{\left(3\right)}a^{-1}\gamma-\frac{1}{3}\gamma_{\left(3\right)}\right)-\frac{4}{3}a^{-1}\gamma_{\left(3\right)}\frac{\rho}{\ell}\mathit{f}\left(\rho\right)\right]d\rho dv\right.\nonumber \\
 & \quad +\frac{2}{3}a^{-1}\left[\left(1-2\mathit{f}\left(\rho\right)\right)^{2}\left(\gamma_{\left(3\right)}\omega-3\gamma\omega_{\left(3\right)}\right)+4a^{-1}\gamma_{\left(3\right)}\omega\frac{\rho}{\ell}\mathit{f}\left(\rho\right)\right]d\rho d\varphi\nonumber \\
 & \quad +\frac{2}{3}a^{-1}\ell\rho\mathit{f}\left(\rho\right)\left[\left(1-2\mathit{f}\left(\rho\right)\right)^{2}\left(\gamma_{\left(3\right)}\left(a^{2}+4a_{\left(3\right)}^{2}\right)-6\mathit{a}a_{\left(3\right)}\gamma\right)+4a\gamma_{\left(3\right)}\frac{\rho}{\ell}\mathit{f}\left(\rho\right)\right]dt^{2}\nonumber \\
 & \quad +\frac{4}{3}a^{-1}\rho\mathit{f}\left(\rho\right)\left[\left(1-2\mathit{f}\left(\rho\right)\right)^{2}\left(3\gamma\left(\mathit{a}\omega_{\left(3\right)}+a_{\left(3\right)}\omega\right)-\gamma_{\left(3\right)}\left(a\omega+4a_{\left(3\right)}\omega_{\left(3\right)}\right)\right)\right.\label{spin3field-NullC}\\
 & \quad +\left.4\omega\gamma_{\left(3\right)}\frac{\rho}{\ell}\mathit{f}\left(\rho\right)\right]dvd\varphi+\left[\left(1-2\mathit{f}\left(\rho\right)\right)^{4}\gamma_{\left(3\right)}\left(\gamma^{2}-\frac{4}{3}\gamma_{\left(3\right)}^{2}\right)\right.\nonumber \\
 & \quad +\frac{1}{3}\left(1-2\mathit{f}\left(\rho\right)\right)^{2}\left(2a^{-1}\left(\gamma_{\left(3\right)}\left(\omega^{2}+4\omega_{\left(3\right)}^{2}\right)-6\gamma\omega_{\left(3\right)}\omega\right)\frac{\rho}{\ell}\mathit{f}\left(\rho\right)-\gamma_{\left(3\right)}\gamma^{2}\right)\nonumber \\
 & \quad \left.\left.+\frac{8}{3}a^{-2}\gamma_{\left(3\right)}\left(\frac{\rho}{\ell}\mathit{f}\left(\rho\right)\omega^{2}+\mathit{a}\gamma_{\left(3\right)}^{2}\right)\frac{\rho}{\ell}\mathit{f}\left(\rho\right)+\frac{28}{27}\gamma_{\left(3\right)}^{3}\right]d\varphi^{2}\right\} d\varphi\;.\nonumber 
\end{align}
Therefore, close to the horizon, the component $\Phi_{\varphi\varphi\varphi}$
expands as in eq. \eqref{phi-Null}.

For a generic choice of Lagrange multipliers, the spacetime geometry
can be described in normal coordinates as in eq. \eqref{Metric-NormalC},
and the exact expression for the spin-3 field is given by

\begin{align}
\Phi & =-\frac{\ell^{3}}{3}\left(\frac{1}{2}\left(\zeta_{\left(3\right)}^{+}-\zeta_{\left(3\right)}^{-}\right)dt+\frac{\gamma_{\left(3\right)}}{\ell}d\varphi\right)dr^{2}\nonumber \\
 &\quad -\frac{\ell^{3}}{6}\left\{ -\frac{3}{8}\left(\cosh\left(4r\right)+\frac{1}{3}\right)\left[\frac{4}{3}\zeta_{\left(3\right)}^{+}\zeta_{\left(3\right)}^{-}\left(\zeta_{\left(3\right)}^{+}-\zeta_{\left(3\right)}^{-}\right)+\zeta_{\left(3\right)}^{+}\left(\zeta^{-}\right)^{2}-\zeta_{\left(3\right)}^{-}\left(\zeta^{+}\right)^{2}\right]\right.\nonumber \\
 &\quad \left.+\zeta^{+}\zeta^{-}\left(\zeta_{\left(3\right)}^{+}-\zeta_{\left(3\right)}^{-}\right)\cosh\left(2r\right)+\frac{2}{9}\left[\left(\zeta_{\left(3\right)}^{+}\right)^{3}-\left(\zeta_{\left(3\right)}^{-}\right)^{3}\right]-\frac{1}{2}\left[\left(\zeta^{+}\right)^{3}-\left(\zeta^{-}\right)^{3}\right]\right\} dt^{3}\nonumber \\
 &\quad -\frac{1}{6}\left\{ \gamma_{\left(3\right)}^{3}\left(\cosh\left(4r\right)+\frac{7}{9}\right)+\omega_{\left(3\right)}\left(3\gamma\omega-2\gamma_{\left(3\right)}\omega_{\left(3\right)}\right)\sinh^{2}\left(2r\right)\right.\nonumber \\
 &\quad \left.-3\gamma_{\left(3\right)}\omega^{2}\sinh^{2}\left(r\right)\left(\cosh\left(2r\right)-\frac{1}{3}\right)-3\gamma_{\left(3\right)}\gamma^{2}\cosh^{2}\left(r\right)\left(\cosh\left(2r\right)+\frac{1}{3}\right)\right\} d\varphi^{3}\nonumber \\
 &\quad -\ell\left\{ \left(\zeta^{+}+\zeta^{-}\right)\left(\gamma\omega_{\left(3\right)}\cosh^{2}\left(r\right)+\frac{1}{6}\gamma_{\left(3\right)}\omega\left(1-3\cosh\left(2r\right)\right)\right)\sinh^{2}\left(r\right)\right.\nonumber \\
 &\quad +\left(\zeta^{+}-\zeta^{-}\right)\left(\omega_{\left(3\right)}\omega\sinh^{2}\left(r\right)-\frac{1}{6}\gamma_{\left(3\right)}\gamma\left(3\cosh\left(2r\right)+1\right)\right)\cosh^{2}\left(r\right)\label{eq:PhiNC}\\
 &\quad +\frac{1}{4}\left(\zeta_{\left(3\right)}^{+}+\zeta_{\left(3\right)}^{-}\right)\left(\gamma\omega-\frac{4}{3}\gamma_{\left(3\right)}\omega_{\left(3\right)}\right)\sinh^{2}\left(2r\right)-\frac{1}{4}\left(\zeta_{\left(3\right)}^{+}-\zeta_{\left(3\right)}^{-}\right)\left[\frac{2}{3}\omega_{\left(3\right)}^{2}\sinh^{2}\left(2r\right)\right.\nonumber \\
 &\quad -\gamma_{\left(3\right)}^{2}\left(\cosh\left(4r\right)+\frac{7}{9}\right)+\gamma^{2}\cosh^{2}\left(r\right)\left(\cosh\left(2r\right)+\frac{1}{3}\right)\nonumber \\
 &\quad \left.\left.+\omega^{2}\sinh^{2}\left(r\right)\left(\cosh\left(2r\right)-\frac{1}{3}\right)\right]\right\} dtd\varphi^{2}-\frac{\ell^{2}}{2}\left\{ \omega_{\left(3\right)}\left[\left(\zeta^{+}\right)^{2}-\left(\zeta^{-}\right)^{2}\right]\sinh^{2}\left(r\right)\cosh^{2}\left(r\right)\right.\nonumber \\
 &\quad -\frac{1}{6}\gamma_{\left(3\right)}\left[\left(\zeta_{\left(3\right)}^{+}+\zeta_{\left(3\right)}^{-}\right)^{2}\sinh^{2}\left(2r\right)-\frac{3}{2}\left(\cosh\left(4r\right)+\frac{7}{9}\right)\left(\zeta_{\left(3\right)}^{+}-\zeta_{\left(3\right)}^{-}\right)^{2}\right.\nonumber \\
 &\quad \left.+\frac{3}{2}\left(\zeta^{+}+\zeta^{-}\right)^{2}\sinh^{2}\left(r\right)\left(\cosh\left(2r\right)-\frac{1}{3}\right)+\frac{3}{2}\left(\zeta^{+}-\zeta^{-}\right)^{2}\cosh^{2}\left(r\right)\left(\cosh\left(2r\right)+\frac{1}{3}\right)\right]\nonumber \\
 &\quad +3\left(\zeta_{\left(3\right)}^{+}-\zeta_{\left(3\right)}^{-}\right)\left[\left(\zeta^{+}-\zeta^{-}\right)\gamma\left(\cosh\left(2r\right)+\frac{1}{3}\right)\cosh^{2}\left(r\right)+\left(\zeta^{+}+\zeta^{-}\right)\omega\left(\cosh\left(2r\right)-\frac{1}{3}\right)\sinh^{2}\left(r\right)\right]\nonumber \\
 &\quad \left.-\frac{3}{2}\left(\zeta_{\left(3\right)}^{+}+\zeta_{\left(3\right)}^{-}\right)\left[\left(\zeta^{+}+\zeta^{-}\right)\gamma+\left(\zeta^{+}-\zeta^{-}\right)\omega-\frac{4}{3}\left(\zeta_{\left(3\right)}^{+}-\zeta_{\left(3\right)}^{-}\right)\omega_{\left(3\right)}\right]\sinh^{2}\left(2r\right)\right\} d\varphi dt^{2}\;,\nonumber 
\end{align}
so that the component $\Phi_{\varphi\varphi\varphi}$ expands according to \eqref{phi-Null}, with ${\cal O}(\rho)$ replaced by ${\cal O}(r^2)$, near the horizon.

\providecommand{\href}[2]{#2}\begingroup\raggedright\endgroup

\end{document}